\documentclass[amssymb,amsmath,aps,groupedaddress,reprint,superscriptaddress]{revtex4}

\usepackage{amsmath}
\usepackage{fixmath}

\usepackage{graphicx}
\usepackage{wrapfig}
\usepackage[toc,page]{appendix}
\graphicspath{{images/}}

\newcommand{\mb}{\mathbold}
\newcommand{\mc}{\mathcal}
\newcommand{\be}{\begin{equation}}
\newcommand{\ee}{\end{equation}}

\begin{document}

\title{Magneto-polaritons in Weyl semimetals in a strong magnetic field}
\author{Zhongqu Long}
\affiliation{Department of Physics and Astronomy, Texas A\&M
University, College Station, TX, 77843 USA}
\author{Yongrui Wang}
\affiliation{Department of Physics and Astronomy, Texas A\&M
University, College Station, TX, 77843 USA}
\author{Maria Erukhimova}
\affiliation{Institute of Applied Physics, Russian Academy of Sciences}
\author{Mikhail Tokman}
\affiliation{Institute of Applied Physics, Russian Academy of Sciences}
\author{Alexey Belyanin}
\affiliation{Department of Physics and Astronomy, Texas A\&M
University, College Station, TX, 77843 USA}

\begin{abstract}

Exotic topological and transport properties of Weyl semimetals generated a lot of excitement in the condensed matter community. Here we show that Weyl semimetals in a strong magnetic field are highly unusual optical materials. The propagation of electromagnetic waves is affected by an interplay between plasmonic response of chiral Weyl fermions and extreme anisotropy induced by a magnetic field. The resulting magneto-polaritons possess a number of  peculiar properties, such as hyperbolic dispersion, photonic stop bands, coupling-induced transparency, and broadband polarization conversion. These effects can be used for optical spectroscopy of these materials including detection of the chiral anomaly, or for broadband terahertz/infrared applications.

\end{abstract} 

\date{\today}

\maketitle

\section{Introduction}

Weyl fermions were studied extensively in the context of relativistic quantum field theory. Recently they have emerged as low-energy excitations in some crystalline solids, named, unsurprisingly, Weyl semimetals (WSMs). The Brillouin zone of WSMs contains pairs of Weyl nodes of opposite chirality, described by the Weyl Hamiltonian,
\be \label{1} 
H = \chi \hbar v_F \mb{\sigma} \mb{k}.  
\end{equation}
Here $\chi = \pm 1$ is the chirality index, $\mb{\sigma}$ is a 3D vector of Pauli matrices, $\mb{k}$  is the 3D quasimomentum of electrons counted from the Weyl node, and we assumed an isotropic electron dispersion (scalar constant $v_F$) for simplicity. 

WSMs have unusual electronic and transport properties originating from the nontrivial topology of the Brillouin zone \cite{wan2011,hosur2013,vafek2014}. They have been studied experimentally, mostly with angle-resolved photoemission spectroscopy; e.g.~\cite{xu2015, lv2015, xu2015-2}. The most intensely studied phenomena include topologically protected surface states known as Fermi arcs, the chiral anomaly, or the non-conservation of the chiral charge in parallel electric and magnetic fields, and the resulting anomalous magnetoresistance \cite{xiong2015,huang2015,son2013}. Optics of WSMs received relatively less attention so far. Far-infrared optical spectroscopy studies of TaAs without the magnetic field have been recently reported \cite{xu2016}. The static and high-frequency conductivity, magnetoplasmons, and polaritons in a magnetic field were calculated recently in classical approximation \cite{spivak2016, zhou2015,pell2015,tabert2016,ashby2014,ma2015}. The strong-field optical conductivity was calculated in \cite{ashby2013}. Here we concentrate on the wave propagation in WSMs in a strong magnetic field, when the electron motion transverse to the field becomes quantized. We show that hybridization of magnetoplasmons with electromagnetic (EM) waves in WSMs leads to a treasure trove of fascinating optical phenomena involving magnetopolaritons: hyperbolic dispersion, the absence of Landau damping for strongly localized excitations, photonic stop bands, coupling-induced transparency, efficient polarization conversion, and pulse compression, to name a few. We show that optical spectroscopic techniques provide a straightforward and ``clean'' way of detecting topological properties of low-energy electron states and in particular the chiral anomaly. Moreover, WSMs show strong promise for future photonic chips enabling a wide array of broadband optoelectronic applications, such as polarizers, modulators, switches, and pulse shapers for mid-infrared through terahertz wavelengths. 

\section{Dielectric tensor for a WSM in a magnetic field}

We consider the material which has only one pair of isotropic Weyl nodes for simplicity. The treatment can be  generalized to an arbitrary number of Weyl point pairs and anisotropic electron dispersion. In a strong magnetic field oriented along $z$ axis the 3D conical spectrum of quasiparticles near each node is split into Landau-level (LL) subbands $W_n$ labeled by the quantum number $n$:
\begin{align}
W_n&=\text{sgn}(n) \hbar v_F \sqrt{\frac{2|n|}{l_b^2}+k_z^2} \text{ for
     } n\neq 0, \\
W_0^{(\chi)} &=-\chi \hbar v_F k_z
\label{energy}
\end{align}
  where  $l_b=\sqrt{\frac{\hbar c}{eB}}$ is the magnetic length. The electron wavefunctions are given in the Supplement  \cite{SM}. 
  
  The salient feature of the electron spectrum is the presence of chiral electron states with 1D linear dispersion at $n = 0$ LL. The $n = 0$ electrons near each node are able to move only in one direction, depending on the sign of $\chi$ and neglecting internode scattering.  The majority of peculiar optical properties of WSMs originates from the response of these electron states and its interplay with inter-LL transitions. 
  
  The  dielectric tensor for chiral fermions in WSMs has a general structure typical for a magnetized electron-hole plasma: 
  \begin{equation}
\epsilon_{ij}= \left(\begin{array}{ccc}
\epsilon_{\perp} & ig & 0 \\
-ig & \epsilon_{\perp} & 0  \\
0 & 0 & \epsilon_{zz}
\end{array}\right)
\label{eij}
\end{equation} 
where $i,j = x,y,z$. However, the expressions for its components and the resulting optical response are far from typical. Consider first the longitudinal component $\epsilon_{zz} = \epsilon_b + 4 \pi i \sigma_{zz}/\omega$ where $\epsilon_b$ is the background dielectric constant of a crystal.  The conductivity $\sigma_{zz}$ can be found by calculating the linear response to the longitudinal field $E_z = {\rm Re}[\mc{E} e^{iq_z z - i \omega t}]$.  It is convenient to define the optical field through the scalar potential $\phi = {\rm Re}[\Phi e^{iq_z z - i \omega t}]$ as $\mc{E} = -i q_z \Phi$. To avoid cumbersome summation over the Landau levels, we will assume for simplicity that the Fermi energy for each chirality is between $n = -1$ and $n = 1$ and the temperature is low enough so that the states with $n \neq 0$ are either completely filled or empty. The general result for an arbitrary position of the Fermi level is given in \cite{SM}. Note also that for the longitudinal field $\mb{E} \| \mb{B}$ the  transitions between the Landau levels are forbidden in the electric dipole approximation. The resulting linearized density matrix equation for the density matrix elements $\rho_{kk'}^{(\chi)}$ for each chirality is 
\begin{align} 
& -i\omega \rho_{k_z,k'_z}^{(\chi)} + i \frac{W_{0}^{(\chi)}(k_z) - W_{0}^{(\chi)}(k'_z)}{\hbar} \rho_{k_z,k'_z}^{(\chi)} =  
 - \frac{i}{\hbar} e \Phi  \left< n = 0,k_z | e^{iq_z z} | n = 0,k'_z \right> \left[ f_{0}^{(\chi)}(k_z) -  f_{0}^{(\chi)}(k'_z) \right], 
\label{dm1}
\end{align}
where $f_{0}^{(\chi)}(k_z)$ are populations at $n = 0$ unperturbed by the optical field and we neglected relaxation, which will be added later. The matrix element in Eq.~(\ref{dm1}) is calculated using the electron states in a magnetic field \cite{SM}; it is equal to the delta-function $\delta_{k_z - q_z,k'_z}$. The solution of Eq.~(\ref{dm1}) in the limit $k_z \gg q_z$ is 
\be 
  \rho_{k_z,k_z - q_z}^{(\chi)} = \frac{ie\mc{E}}{\omega - \chi q_z v_F} \frac{\partial f_{0}^{(\chi)}(k_z)}{\hbar \partial k_z}.
  \label{dm2}
 \ee
 The complex amplitude of the Fourier component of the electric current $j_z = {\rm Re}\tilde{J} e^{iq_z z - i \omega t}$ is given by 
 \be 
 \tilde{J} = \sum_{k_z,\chi} \left( j_z \right)_{k_z - q_z,k_z}^{(\chi)}  \rho_{k_z,k_z - q_z}^{(\chi)},
 \label{j1}
 \ee
 where the matrix element of the spatial Fourier component of the current is 
 \be
 \left( j_z \right)_{k'_z,k_z}^{(\chi)} = -e \left< n = 0,k'_z | e^{-i q_z z} \chi v_F \sigma_z | n = 0,k_z \right>
 \label{j2} 
 \ee
 and the sum can be replaced by integration. The resulting longitudinal component of the conductivity tensor is 
 \be
 \sigma_{zz} = \frac{2 i e^3 B v_F \omega}{4 \pi^2 \hbar^2 c} \frac{1}{\omega^2 - q_z^2 v_F^2},
 \label{cond}
 \ee
 where the $B$-dependence appeared due to the density of states in a quantizing magnetic field. The longitudinal dielectric tensor component therefore takes the form 
  \be
 \epsilon_{zz} = \epsilon_b -   \frac{\omega_p^2}{\omega^2 - q_z^2 v_F^2}.
 \label{ezz}
 \ee
 
This result can be also obtained from the kinetic equation \cite{SM}. This expression has several peculiar features.  First, since the electrons at $n = 0$ can move only in one direction with the same velocity $v_F$, they cannot bunch in the velocity space and there is no Landau damping. Mathematically, the Landau damping emerges due to contribution from the pole in the integral over electron momenta in the linear conductivity. However, in our case there is no pole in the integral in Eq.~(\ref{j1}) since the denominator in Eq.~(\ref{dm2}) does not depend on the electron momentum. 

Second, the effective plasma frequency in Eq.~(\ref{ezz}) does not depend on the electron density at all and instead depends on the magnetic field: $\omega_p^2 = \displaystyle   \frac{2\alpha}{\pi} \frac{eBv_F}{\hbar}$, where $\alpha = \displaystyle \frac{e^2}{\hbar c}$; see also \cite{son2013,spivak2016}. In the limit of a uniform electric field $q_z = 0$ Eq.~(\ref{cond}) immediately gives the chiral anomaly. Indeed, if only $n = 0$ electrons are involved, the chiral current $j_{\rm chir} = \displaystyle \frac{\partial [N^{(\chi = +1)} -N^{(\chi = -1)}]}{\partial t}$ is related to the charge current in a uniform but time-dependent field $\mb{E} \| \mb{B}$ as $\partial j_z/\partial t = - e v_F j_{\rm chir}$. This gives the chiral anomaly current $j_{\rm chir} = \displaystyle - \frac{e^2 \mb{E} \mb{B}}{2 \pi^2 \hbar^2 c}$, in agreement with previous results; see e.g.~\cite{hosur2013,vafek2014} for review.

 \section{Properties of magneto-polaritons}

\subsection{The dispersion equation}

EM waves incident on a magnetized WSM propagate as eigenmodes that can be called magneto-polaritons. They are the solutions of Maxwell's equations for plane waves,
\be
\sum_j q_i q_j E_j - q^2 E_i  + \frac{\omega^2}{c^2} \epsilon_{ij} E_j = 0
\label{fresnel}
\ee
with the dielectric tensor from Eq.~(\ref{eij}). 
For the photon wave vector $\mb{q}$ in the $(xz)$-plane making an angle $\theta$ with the magnetic field direction along $z$-axis, Eqs.~(\ref{fresnel}) can be written as 
\begin{widetext}
   \begin{equation}
 \left(\begin{array}{ccc}
\epsilon_{+} -\frac{1}{2} n^2 \left( 1+ \cos^2\theta \right) & \frac{1}{2} n^2 \sin^2\theta & \frac{1}{\sqrt{2}} n^2 \sin\theta \cos\theta \\
\frac{1}{2} n^2 \sin^2\theta  &\epsilon_{-} -\frac{1}{2}  n^2  \left( 1+ \cos^2\theta \right) & \frac{1}{\sqrt{2}} n^2 \sin\theta \cos\theta   \\
\frac{1}{\sqrt{2}} n^2 \sin\theta \cos\theta  & \frac{1}{\sqrt{2}} n^2 \sin\theta \cos\theta  & \epsilon_{zz} - n^2 \sin^2\theta
\end{array}\right) 
 \left(\begin{array}{c} E_+ \\ E_- \\ E_z \end{array}\right) = 0, 
\label{fresnel2}
\end{equation} 
\end{widetext}
where $n^2 = \displaystyle \frac{c^2 q^2}{\omega^2}$, $\epsilon_{\pm} = \epsilon_{\perp} \pm g$, and $E_{\pm} = \frac{1}{\sqrt{2}} (E_x \pm i E_y)$. \\

\subsection{Longitudinal propagation} 

For the waves propagating strictly along the magnetic field, i.e.~$\theta = 0$, the solution to Eqs.~(\ref{fresnel2}) consists of two eigenmodes with transverse polarization (``photons''), 
\be
n_{L,R}^2 = \epsilon_{\pm}, \; \mb{E_{L,R}} = \frac{1}{\sqrt{2}} E_{\pm} (\mb{x_0} \pm i \mb{y_0} ),
\label{LR}
\ee
and the wave with the longitudinal polarization $\mb{E} = E_z \mb{z_0}$ and dispersion equation $\epsilon_{zz} = 0$ (``plasmon''). The plasmon dispersion is 
\be
\omega^2 = \frac{\omega_p^2}{\epsilon_b} + v_F^2 q^2.
\label{plasmon}
\ee 
We emphasize again that, in contrast to ``usual'' plasmons, there is no cutoff in Eq.~(\ref{plasmon}) due to Landau damping at large wave vectors $q > \omega/v_F $. Therefore, a much stronger plasmon localization is allowed, with propagation only limited by absorption due to scattering on impurities etc. 

\subsection{Oblique propagation}

For oblique propagation, even at very small angles $\theta$, the plasmons and transverse waves are coupled to form hybrid plasmon-polaritons. To determine general trends and obtain analytic results, we neglect the spatial dispersion ($q_z$-dependence) of $\epsilon_{zz}$ in Eq.~(\ref{ezz}), which is possible as long as $n^2 \sin^2\theta \ll c^2/v_F^2$. This is not so restrictive since $c/v_F > 100$. We also neglect any spatial dispersion in $\epsilon_{\pm}$ in the dipole approximation. 

 It is instructive first to consider the case when the Fermi level is exactly at the Weyl point for both chiralities. Some of the unique features of wave propagation in WSMs manifest themselves already in this limit.  In this case, due to electron-hole symmetry the off-diagonal terms in Eq.~(\ref{eij}) vanish and the dielectric tensor looks like the one for a uniaxial anisotropic medium. The dispersion equation for the extraordinary wave, i.e.~the one polarized in the ($x,z$) plane, can be written as 
\be 
\label{hyper}
\frac{n_x^2}{\epsilon_{zz}} + \frac{n_z^2}{\epsilon_{\perp}} = 1.
\ee
The transverse components of the dielectric tensor are always positive, whereas the $\epsilon_{zz}$ component becomes negative for frequencies below the plasmon resonance, $\omega^2 < \omega_p^2/\epsilon_b$. In this case  Eq.~(\ref{hyper}) becomes hyperbolic and its isofrequency lines are hyperbolae. Therefore, a magnetized WSM is a natural hyperbolic material at low enough frequencies. Another natural hyperbolic material is hexagonal boron nitride, where the hyperbolic dispersion exists in two narrow spectral ranges near the phonon bands \cite{hbn}. Otherwise, hyperbolic dispersion is achieved only in the effective medium approximation in metal/dielectric metamaterials prepared by nanofabrication \cite{zayats}. It is promising for numerous applications from superlenses and nanoimaging to photonic integrated circuits. The plasmon resonance frequency $\omega = \omega_p/\sqrt{\epsilon_b}$ in WSMs which determines the upper bound for hyperbolic dispersion is in the THz to far-infrared range for a magnetic field of 1-10 Tesla, $\epsilon_b \sim 10$ and $v_F \sim 10^8$ cm/s. It is lower than the inter-LL absorption edge for all magnetic fields, so the only loss mechanism is due to scattering on impurities which depends on the material quality. 

The ordinary wave in this limit is linearly polarized along $y$-axis and has a standard dispersion $n^2 = \epsilon_{\perp}$. 

\begin{figure}[h]
    \centering
    \includegraphics[width=0.5\columnwidth]{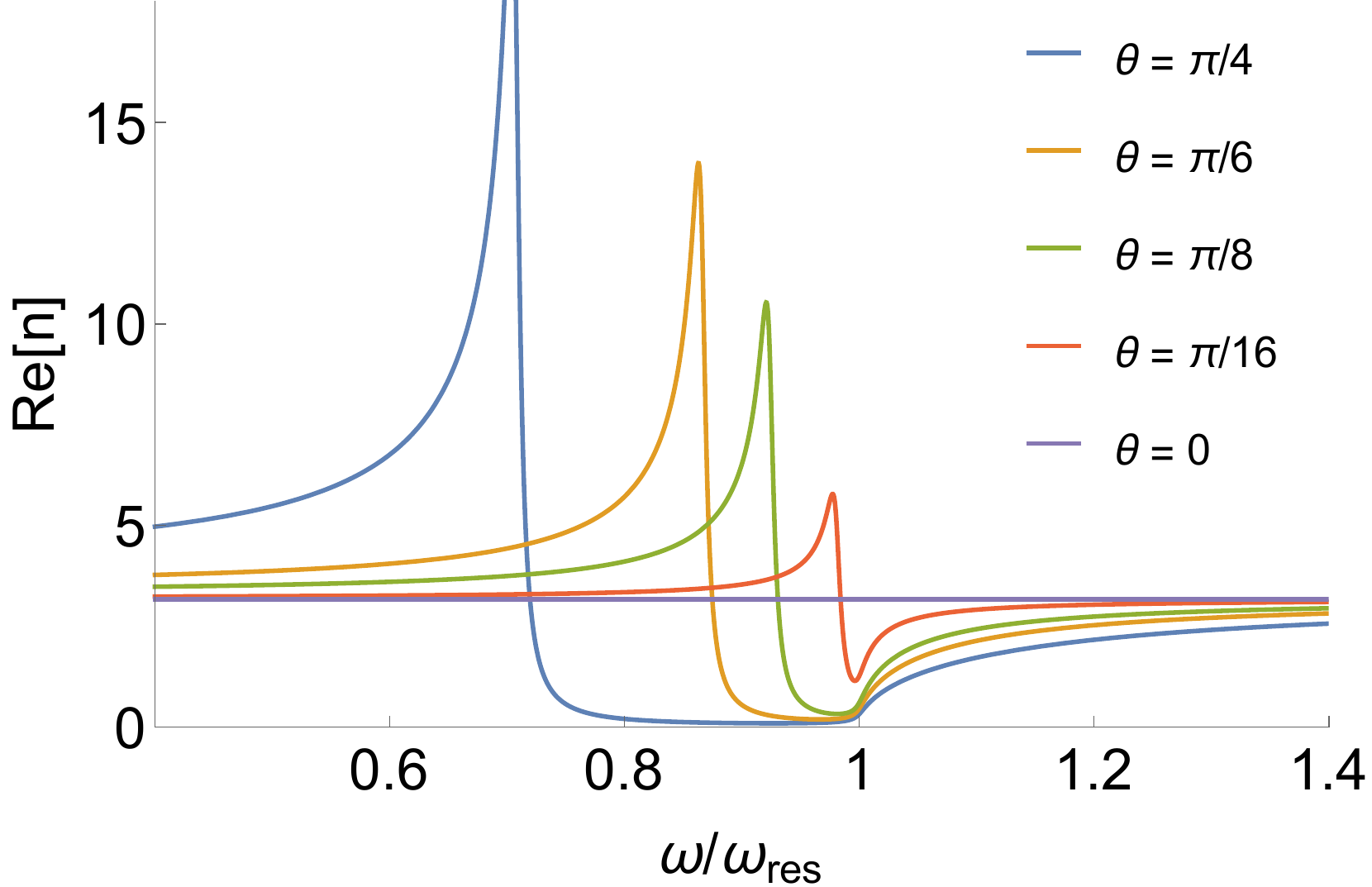}
    \caption{Dispersion (real part of $n(\omega)$) of the extraordinary waves in a magnetic field of 10 T for several different propagation angles $\theta$.}
	\label{polaritons}
\end{figure}

Fig.~\ref{polaritons} shows the dispersion (real part of $n$) for the extraordinary waves for several different propagation angles $\theta$. Far from inter-LL transitions, we can neglect any dispersion in the transverse part of the dielectric tensor, assuming $\epsilon_{\perp} = \epsilon_b \sim 10$. 
We also added the scattering rate as an imaginary part of frequency ($\omega + i \gamma$) in the first term of Eq.~(\ref{dm1}) and took $\gamma$ to be 0.01 of the plasmon resonance frequency $\omega_{\rm res} = \omega_p/\epsilon_b^{1/2}$.  
For longitudinal propagation $\theta = 0$ the photon dispersion is trivial: $n = \sqrt{\epsilon_b}$. For any nonzero angle, plasmons and photons hybridize. At the hybrid plasmon-polariton resonance $n$ diverges in the absence of dissipation. The stop band appears between the hybrid resonance and plasmon resonance, in which $n^2 < 0$ so the field is evanescent.  Note that at the boundaries of the stop band Re[$n$] goes through the value of 1 with a large derivative, leading to a small group velocity $v_{gr} \ll c$. This means that a layer of  WSM is able to compress a pulse incident from vacuum by a factor $c/v_{gr}$.  All spectral features are tunable by varying the magnetic field or the propagation angle $\theta$. 

Going back to the general case of an arbitrary Fermi level, Eqs.~(\ref{fresnel2}) lead to a biquadratic dispersion equation for $n$: 
\begin{align} 
& \frac{n^4}{2} \left( (\epsilon_+ + \epsilon_-) \sin^2\theta + 2 \epsilon_{zz} \cos^2\theta \right)  -  
 \left( \epsilon_+ \epsilon_- \sin^2\theta \,  +   \frac{1}{2} \epsilon_{zz} (\epsilon_+ + \epsilon_-) (1 + \cos^2\theta) \right) n^2 + \epsilon_+ \epsilon_- \epsilon_{zz} = 0. 
\label{fresnel3}
\end{align}
The polarization coefficients of the normal modes are 
\be
\label{pol}
\frac{E_{\pm}}{E_z} = \frac{-\frac{1}{\sqrt{2}} n^2 \sin\theta \cos\theta (\epsilon_{\mp} - n^2)}{\epsilon_+ \epsilon_- - \frac{1}{2} n^2 (1+\cos^2\theta)(\epsilon_+ + \epsilon_-) + n^4 \cos^2\theta}.
\ee

Equations (\ref{fresnel3}), (\ref{pol}), and (\ref{ezz}) provide a complete analytic description of the electromagnetic wave propagation in WSMs. They can be plotted numerically or solved analytically, leading to cumbersome formulas. In the low temperature limit we obtain analytic expressions for all components of the dielectric tensor, see \cite{SM}.  Leaving detailed numerical studies to future publications, here we highlight the most interesting cases that can be easily addressed analytically. 

\subsection{Coupling-induced transparency} 

For quasi-longitudinal propagation, $\sin^2\theta \ll 1$ and plasmon-polariton hybridization occurs in the vicinity of the plasmon resonance, $|\epsilon_{zz}| \ll 1$. In this case the approximate solution of Eq.~(\ref{fresnel3}) is 
\begin{align} 
& n_{1,2}^2 = \frac{1}{ (\epsilon_+ + \epsilon_-) \sin^2\theta + 2 \epsilon_{zz}} \left[ \epsilon_+ \epsilon_- \sin^2\theta + \epsilon_{zz} (\epsilon_+ + \epsilon_-) \pm \sqrt{( \epsilon_+ \epsilon_- \sin^2\theta)^2 + \epsilon_{zz}^2  (\epsilon_+ - \epsilon_-)^2 } \right].
\label{hybrid1}
\end{align}
The polarization coefficients become 
\be \label{pol2}
K_{\pm} = \frac{E_{\pm}}{E_z} =  \frac{-\frac{1}{\sqrt{2}} n^2 \sin\theta }{\epsilon_{\pm} - n^2}.
\ee
In the ``non-gyrotropic'' limit when $E_F = 0$ and $\epsilon_+ = \epsilon_- = \epsilon_{\perp}$, the extraordinary wave has dispersion 
\be
\label{nongyro}
n_{2}^2 = \frac{ \epsilon_{zz} \epsilon_{\perp} }{ \epsilon_{\perp} \sin^2\theta +  \epsilon_{zz}}; \; K_+ = K_- = - \frac{1}{\sqrt{2}} \frac{\epsilon_{zz}}{\epsilon_{\perp} \sin\theta}.
\ee
The hybrid resonance corresponds to the vanishing real part of the denominator for $n_2^2$ in Eq.~(\ref{hybrid1}) or (\ref{nongyro}), when $|n_2^2| \gg 1$. 

The effect of {\it coupling-induced transparency} emerges near the plasmon resonance where $n_2$ can be of the order of 1 or smaller. When the angle $\theta$ is not too small, $|\epsilon_{zz}| \ll 1$,  $\sin^2\theta \ll 1$, but $|\epsilon_{\pm}| \sin^2\theta \gg |\epsilon_{zz}|$, the dispersion and polarization of the ``extraordinary'' wave (the wave that becomes extraordinary if $E_F = 0$) are simply 
\be
\label{hybrid2}
n_2^2 = \frac{\epsilon_{zz}}{\sin^2\theta}; \; \frac{E_{x,y}}{E_z} = - \frac{\epsilon_{zz}}{2\sin\theta} \frac{\epsilon_- \pm  \epsilon_+}{\epsilon_+  \epsilon_-}.
\ee
In this case one can have $|n^2| \ll \epsilon_{\pm}$ whereas the electric field of the wave is directed almost along the magnetic field, i.e.~ still quasi-longitudinal. 
Note that $n_2^2$ in Eq.~(\ref{hybrid2}) depends only on the $\epsilon_{zz}$ component, which means that the propagation is not affected at all by the resonant inter-LL absorption losses described by the imaginary parts of $\epsilon_{\pm}$.  The medium effectively becomes transparent for this wave! More accurately, its losses are determined only by the imaginary part of $\epsilon_{zz}$, i.e.~disorder-related scattering. The emergence of transparency is physically obvious from the fact that the polarization of the wave is almost along $B$, and therefore it is decoupled from the transitions between LLs in the electric dipole approximation. The narrow band of transparency within a broad line of inter-LL absorption is defined by the range of frequencies where $|\epsilon_{zz}|$ is small enough, namely $ |\epsilon_{zz}| \ll |\epsilon_{\pm}| \sin^2\theta $. The situation is similar to the electromagnetically induced transparency (EIT) \cite{EIT}, only in the case of EIT the coupling between two quantum oscillators is provided by a coherent EM drive; see \cite{ET2015} where the cases of hybridization coupling and drive-induced coupling are compared.

 In the same limit the ``ordinary'' wave (or rather the wave that becomes ordinary when $E_F = 0$) has the dispersion $n_1^2 = \displaystyle \frac{2 \epsilon_+ \epsilon_-}{\epsilon_+ + \epsilon_-}$ and elliptical polarization in the plane of vectors $\mb{q}$ and $\mb{y_0}$: $\displaystyle \frac{E_x}{E_z} = \sin\theta$ and $\displaystyle \frac{E_y}{E_z} = \displaystyle  \frac{i (\epsilon_+ + \epsilon_-)}{\epsilon_+ - \epsilon_-} \sin\theta$. 

\begin{figure}[h]
    \centering
    \includegraphics[width=0.5\columnwidth]{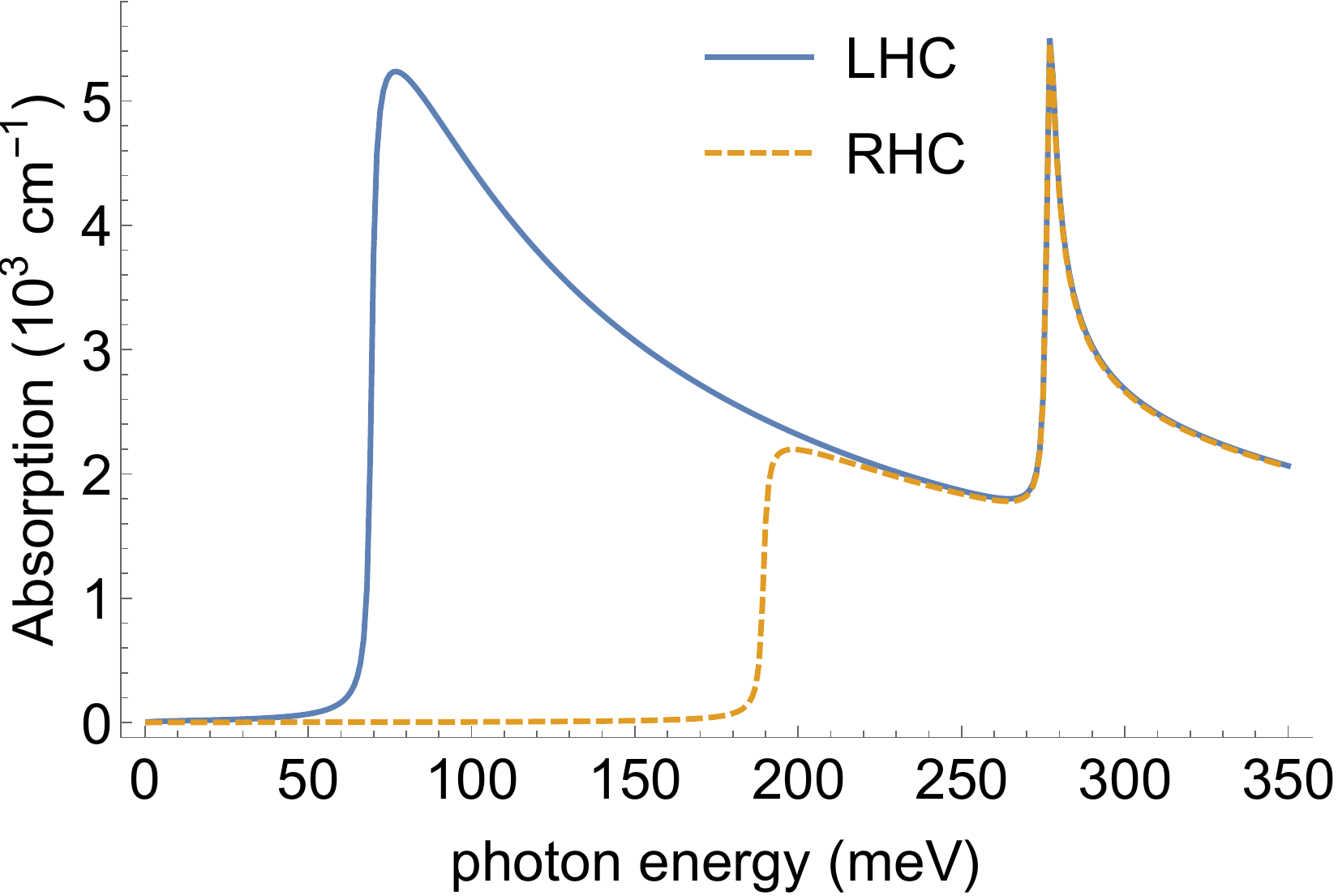}
    \caption{Absorption spectrum for LHC (solid line) and RHC (dashed line) polarizations in a magnetic field $B = 10$ T at zero temperature, the Fermi energy of 60 meV, and the relaxation constant $\gamma = 1$ meV.}
	\label{60mev}
\end{figure}

\subsection{Intersubband transitions and optical detection of the chiral anomaly} 

So far we considered peculiar optical properties of WSMs due to massless 1D chiral fermions at the $n = 0$ LL.  Here we show that resonant inter-LL absorption from $n = 0$ to $n \neq 0$ states provides another clean and sensitive method of studying chiral fermions near Weyl nodes and in particular, detecting the chiral anomaly. Consider the propagation of transverse modes in the Faraday geometry when the eigenmodes are left-hand or right-hand circularly polarized (LHC or RHC). The derivation of the conductivity is straightforward and is outlined in \cite{SM}. Fig.~\ref{60mev}  gives an example of the absorption spectrum at low temperatures when the Fermi level $E_F = 60$ meV is between $ n = 0$ and $n = 1$ LLs and has the same value for both chiralities. Absorption edges  of the lowest-energy transitions $0 \rightarrow 1$, then $-1 \rightarrow 0$, $-2 \rightarrow 1$, and $-1 \rightarrow 2$ are clearly visible in different polarizations (the last two transitions coincide). In particular, there is a broad range of frequencies between 50 and 200 meV when only the LHC polarization is absorbed. Therefore, a several $\mu$m thin WSM film can serve as a broadband polarizer converting from linear into circular polarization. Note that both the frequency bandwidth and the polarization coefficient are tunable by a magnetic field and/or Fermi level position. Other obvious applications include optical isolators and saturable absorbers. 

\begin{figure}[h]
    \centering
    \includegraphics[width=0.5\columnwidth]{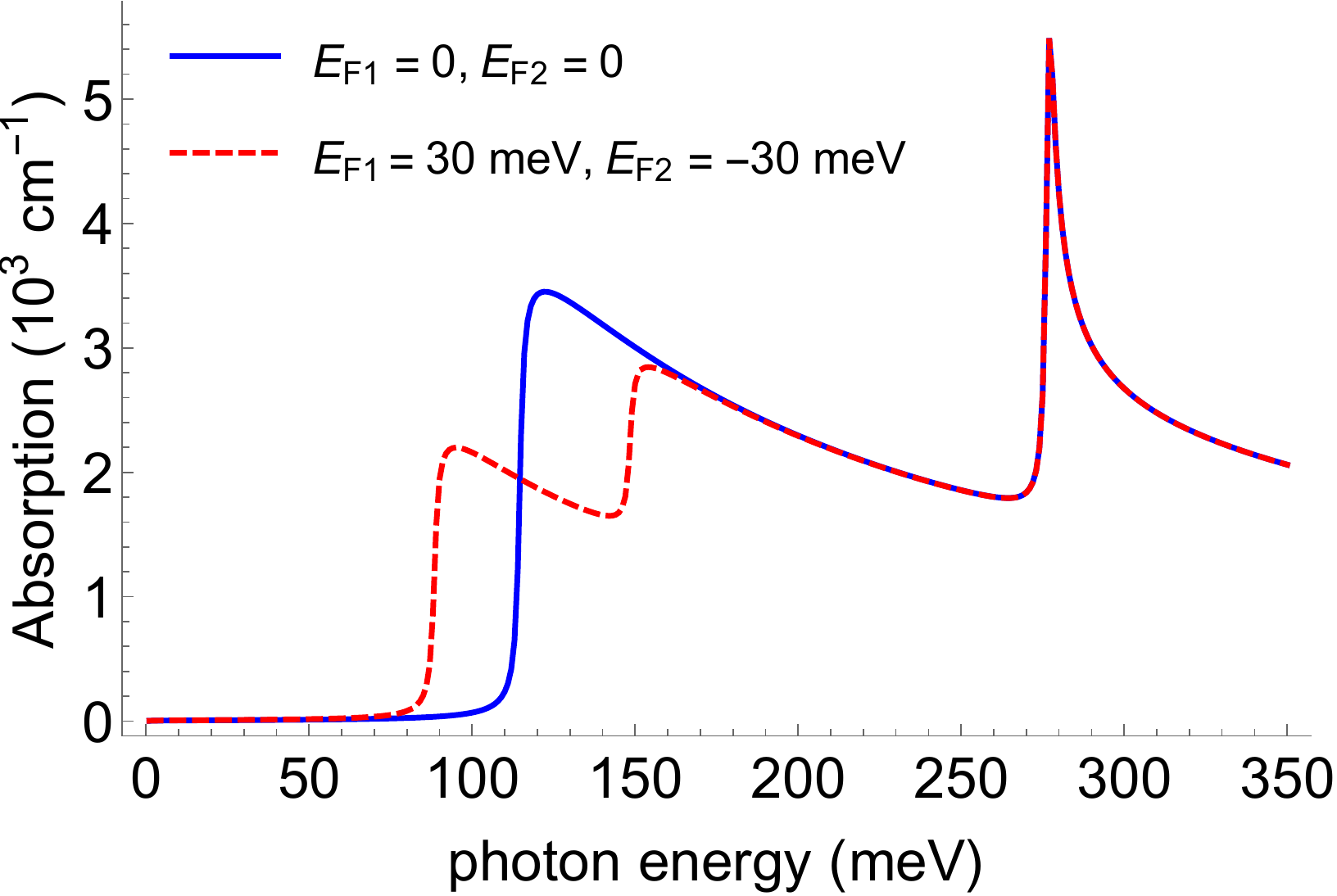}
    \caption{Absorption spectrum before (solid line) and after (dashed line) a constant electric field $\mb{E} \| \mb{B}$ is applied which shifts the Fermi levels by $\pm 30$ meV in the two Weyl nodes. The magnetic field is 10 T and the relaxation constant $\gamma = 1$ meV. }
	\label{anomaly}
\end{figure}

Fig.~\ref{anomaly} shows the evolution of this spectrum when a constant voltage is applied parallel to the magnetic field, which shifts the Fermi levels for the two chiralities by $\pm 30$ meV. Here we assumed that before applying bias, the material was undoped and the Fermi energy was at the Weyl points. In this case both polarizations have the same spectrum. However, when a voltage is applied, an additional absorption edge appears in the spectrum for each polarization, which will be clearly distinguishable as long as the magnitude of the Fermi energy shift is larger than $k_B T$. If a WSM was originally doped, similar absorption edges would appear at different spectral positions for the two polarizations.

In conclusion, we showed that unique topological properties of low-energy quasiparticles in WSMs give rise to a plethora of highly unusual magneto-optical effects, which provide an efficient way of studying these fascinating materials and can be utilized in future photonic devices in the terahertz through mid-infrared range. All effects are broadly tunable by varying the magnetic field, electric bias, or the propagation angle. We hope that our study will stimulate further experimental work in this rapidly developing field. 

This work has been supported by the Air Force Office for Scientific Research
through Grant No.~FA9550-15-1-0153. 
M.E.~and M.T.~acknowledge the support by the RFBR grant No.~17-02-00387. 

\section{Supplemental Material}

\subsection{Electron states in a magnetic field}

When a uniform magnetic field is applied to a WSM, the Weyl cones
split into Landau subbands.

The Hamiltonian of a Weyl electron in a magnetic field is 
\begin{equation} 
\mathcal{H} = \chi  v_F \mb{\sigma} \left( \mb{p}+\frac{e}{c}\mb{A} \right)
\end{equation}
Here $\chi =\pm 1$ is chirality index. If we choose the vector potential $\mb{A} = (0, Bx, 0)$, then 
the Hamiltonian is
\begin{equation}
\mathcal{H} = \chi v_F\left( \begin{array}{cc}
                          \hbar k_z&\pi_x-i\pi_y\\
                          \pi_x+i\pi_y&-\hbar k_z
                                        \end{array}
                                        \right)
                   =  \chi \hbar v_F\left( \begin{array}{cc}
                          k_z&\frac{\sqrt{2}}{l_B}a\\
                          \frac{\sqrt{2}}{l_B}a^{\dagger}&-k_z
                                        \end{array}
                                        \right)
\end{equation}
Here $\mb{\pi} = \mb{p}+\frac{e}{c}\mb{A}$, $l_B = \sqrt{\hbar c/eB}$, and we introduced creation and annihilation operators,
\begin{align}
a =\frac{l_B}{\sqrt{2}\hbar}(\pi_x-i\pi_y), \; 
a^{\dagger} =\frac{l_B}{\sqrt{2}\hbar}(\pi_x+i\pi_y), \nonumber 
\end{align}
which satisfy
\begin{align}
a^{\dagger}\psi_{n0}=\sqrt{|n|}\psi_{n1}, \; 
a\psi_{n1}&=\sqrt{|n|}\psi_{n0} \nonumber 
\end{align}
when the eigenstates are sought in the form
\begin{equation}
\Psi_n(k_y,r)=
  \left ( \begin{array}{c}
            u_n\psi_{n0}\\
            v_n\psi_{n1}
  \end{array} \right ). \nonumber 
\end{equation}
Then we can obtain from $\mathcal{H}\Psi_n=W_n^{(\chi)}\Psi_n$:
\begin{equation}
\chi \hbar v_F\left( \begin{array}{cc}
                          k_z&\frac{\sqrt{2}}{l_B}a\\
                          \frac{\sqrt{2}}{l_B}a^{\dagger}&-k_z
                                        \end{array}
                                        \right)
\left ( \begin{array}{c}
            u_n\psi_{n0}\\
            v_n\psi_{n1}
     \end{array} \right )
= \chi  \hbar v_F\left ( \begin{array}{c}
            (k_z u_n+\frac{\sqrt{2|n|}}{l_B}v_n)\psi_{n0}\\
            (-k_z v_n+\frac{\sqrt{2|n|}}{l_B}u_n)\psi_{n1}
          \end{array} \right )
= W_n^{(\chi)}\Psi_n. \nonumber 
\end{equation}
The eigenenergies are
\begin{align}
W_0^{(\chi)}&= - \chi \hbar v_F k_z\\
W_n&=\text{sgn}(n) \hbar v_F \sqrt{\frac{2|n|}{l_b^2}+k_z^2} \text{ for
     } n\neq 0\\
u_n&=\sqrt{\frac{1}{2}(1 + \frac{\hbar v_F k_z}{W_n})}\\
v_n&=\sqrt{\frac{1}{2}(1 - \frac{\hbar v_F k_z}{W_n})}
\end{align}
and the full expression for eigenstates is
\begin{equation}
\label{psi}
\Psi_n(k,r)=\frac{C_n}{\sqrt{L_yL_z}} \text{exp}(-i(k_yy+k_zz))
      \left( 
         \begin{array}{c}
           \text{sgn}(n) i^{|n|-1} \phi_{|n|-1}u_n\\
           i^{|n|}\phi_{|n|}v_n
         \end{array}
      \right).
\end{equation}
Here
\begin{align}
C_n&=\left\{
             \begin{array}{cl}
             1&n=0\\
             \frac{1}{\sqrt{2}}&n\neq 0
             \end{array}
             \right.\\
\label{phi}
\phi_{|n|}&= \displaystyle \frac {\text{H}_{n}(\frac{x-l_B^2 k_y}{l_B})}
            {\sqrt{2^{|n|} |n|! \sqrt{\pi} l_B}}
            \text{exp}\left(-\frac{1}{2}
            \left(\frac{x-l_B^2k_y}{l_B}\right)^2\right).
\end{align}

\subsection{Selection rules for transitions between Landau levels}

The selection rules for 3D chiral fermions in a magnetic field are very similar
to that for 2D electrons in graphene. For a monochromatic optical field at frequency $\omega$ propagating along $z$ direction and described by the vector potential  $\mb{A} = -i(c/\omega)\mb{E}$ in the $xy$ plane, the
interaction Hamiltonian is
\begin{equation}
\mathcal{H}_{\text{int}}=\chi  v_F\mb{\sigma}\frac{e}{c}\mb{A}.
\end{equation}

The probability of an optical transition between state $m$ and $n$ is
determined by the matrix element  $<m|\mathcal{H}_{\text{int}}|n>$
\begin{equation}
  <m|\mathcal{H}_{\text{int}}|n>= -\frac{i\chi  v_Fe}{\omega }<m|\sigma_x\mb{x_0}+\sigma_y\mb{y_0}|n>\mb{E}\label{intH}
\end{equation}
\begin{equation}
  \sigma_x\mb{x_0}+\sigma_y\mb{y_0}=\left( \begin{array}{cc} 
                                                         0&\mb{x_0}-i\mb{y_0}\\
                                                         \mb{x_0}+i\mb{y_0}&0
                                                         \end{array}
                                                       \right)
                                                     =\sqrt{2}\left(\begin{array}{cc}
                                                         0&\mb{e}_{RHC}\\
                                                         \mb{e}_{LHC}&0
                                                          \end{array}\right) \nonumber 
\end{equation}

Writing $\mb{E} = E_L \mb{e}_{LHC} + E_R \mb{e}_{RHC}$ and substituting the wave functions from Eq.~(\ref{phi}) into Eq.~(\ref{intH}), we obtain 
\begin{align}
<m|\mathcal{H}_{\text{int}}|n>&= -\frac{i\chi  v_Fe}{\omega c
                                }C_m C_n v_m u_n i^{-|m|+|n|-1}<\phi_{|m|}|\phi_{|n|-1}>\sqrt{2} E_R \nonumber \\
                                          &-\frac{i\chi  v_Fe}{\omega
                                c}C_n C_m v_n u_m i^{-|m|+|n|+1}<\phi_{|m|-1}|\phi_{|n|}>\sqrt{2} E_L
\end{align}
The resulting selection rules are:
\begin{align}
|m|&=|n|-1 \text{ for } \mb{e}_{RHC} \text{ polarization}\\
|n|&=|m|-1 \text{ for } \mb{e}_{LHC} \text{ polarization}
\end{align}

\subsection{Transverse optical conductivity due to transitions between Landau levels}

The current density generated in response to a monochromatic field is 
\begin{align}
\mb{j}&=\frac{i(e v_F)^2\hbar}{4\pi^2l_b^2}\int\text{d}k_z\sum_{mn}[(\sigma_R)_{nm}\mb{e}_{RHC}+(\sigma_L)_{nm}\mb{e}_{LHC}][(\sigma_R)_{mn}E_L+(\sigma_L)_{mn}E_R]    \nonumber \\
          &\times \frac{f_n(k_z)-f_m(k_z)}{[\hbar\omega-(W_m(k_z)-W_n(k_z)+i\gamma)](W_m(k_z)-W_n(k_z))}
\end{align}
The corresponding conductivities for LHC and RHC polarizations are given by
\begin{equation}
\tilde{\sigma}_{RR}=\frac{i(e v_F)^2\hbar}{4\pi^2l_b^2}\int\text{d}k_z\sum_{mn}[(\sigma_R)_{nm}(\sigma_L)_{mn}]\times \frac{f_n(k_z)-f_m(k_z)}{[\hbar\omega-(W_m(k_z)-W_n(k_z))+i\gamma](W_m(k_z)-W_n(k_z))} \nonumber 
\end{equation}
\begin{equation}
\tilde{\sigma}_{LL}=\frac{i(e v_F)^2\hbar}{4\pi^2l_b^2}\int\text{d}k_z\sum_{mn}[(\sigma_L)_{nm}(\sigma_R)_{mn}]\times \frac{f_n(k_z)-f_m(k_z)}{[\hbar\omega-(W_m(k_z)-W_n(k_z))+i\gamma](W_m(k_z)-W_n(k_z))} \nonumber 
\end{equation}
Here $f_{n,m}(k_z)$ are occupation numbers of electron states; 
\begin{align}
(\sigma_R)_{nm}&=<n|
                       \left( \begin{array}{cc}
                          0&1\\
                          0&0
                                        \end{array}
                                        \right)|m>  \nonumber \\
&=\sqrt{2}C_nC_mv_nu_m\delta(|n|-|m|-1) \nonumber \\
(\sigma_L)_{nm}&=<n| 
                          \left( \begin{array}{cc}
                          0&0\\
                          1&0
                                        \end{array}
                                        \right)|m> \nonumber \\
&=\sqrt{2}C_nC_mv_mu_n\delta(|n|-|m|+1) \nonumber 
\end{align}

The resulting expressions for $\tilde{\sigma}_{RR}$ and $\tilde{\sigma}_{LL}$
are
\begin{align}
(\tilde{\sigma}_{RR})_{n \rightarrow m}&=\frac{i(e
                              v_F)^2\hbar}{2\pi^2l_b^2}\int\text{d}k_z(C_nC_mv_nu_m)^2\frac{f_n-f_m}{\Delta
                              W (\hbar \omega-\Delta W +i\gamma)}\delta(|n|-|m|-1)\\
(\tilde{\sigma}_{LL})_{n \rightarrow m}&=\frac{i(e
                              v_F)^2\hbar}{2\pi^2l_b^2}\int\text{d}k_z(C_nC_mv_mu_n)^2\frac{f_n-f_m}{\Delta
                              W (\hbar \omega-\Delta W +i\gamma)}\delta(|n|-|m|+1),
\end{align}
where  $\Delta W = W_m(k_z)-W_n(k_z)$; the quantities $C$, $v$, $u$ and $W_n$ are defined earlier.

As an example, we provide below an explicit form of the conductivity components for lowest-energy transitions $0 \rightarrow 1$, $-1\rightarrow 0$, $-1\rightarrow 2$ and $1\rightarrow 2$ for an arbitrary chirality $\chi = \pm 1$: 
\begin{align}
(\tilde{\sigma}_{LL})_{0\rightarrow 1}&=\frac{i(e
                              v_F)^2\hbar}{2\pi^2l_b^2}\int\text{d}k_z(C_1C_0v_0u_1)^2\frac{f_n-f_m}{\Delta
                              W (\hbar \omega-\Delta W +i\gamma)} \nonumber \\
&=\frac{i(ev_F)^2\hbar}{2\pi^2l_b^2}\int\text{d}k_z\frac{1}{4}(1+\chi 
  k_z(\frac{2}{
  l_b^2}+k_z^2)^{-0.5})\nonumber  \\
&\times \frac{f_n-f_m}{\hbar
  v_F(\sqrt{\frac{2}{l_b^2}+k_z^2}+\chi  k_z) (\hbar \omega-\hbar
  v_F(\sqrt{\frac{2}{l_b^2}+k_z^2}+\chi  k_z) +i\gamma)} \nonumber \\
(\tilde{\sigma}_{RR})_{-1\rightarrow 0}&=\frac{i(e
                              v_F)^2\hbar}{2\pi^2l_b^2}\int\text{d}k_z(C_{-1}C_0v_{-1}u_0)^2\frac{f_n-f_m}{\Delta
                              W (\hbar \omega-\Delta W +i\gamma)} \nonumber \\
&=\frac{i(ev_F)^2\hbar}{2\pi^2l_b^2}\int\text{d}k_z\frac{1}{4}(1-\chi 
  k_z(\frac{2}{
  l_b^2}+k_z^2)^{-0.5}) \nonumber \\
&\times \frac{f_n-f_m}{\hbar
  v_F(\sqrt{\frac{2}{l_b^2}+k_z^2}-\chi  k_z) (\hbar \omega-\hbar
  v_F(\sqrt{\frac{2}{l_b^2}+k_z^2}-\chi  k_z) +i\gamma)} \nonumber \\
(\tilde{\sigma}_{LL})_{-1\rightarrow 2}&=\frac{i(e
                              v_F)^2\hbar}{2\pi^2l_b^2}\int\text{d}k_z(C_{-1}C_2v_{-1}u_2)^2\frac{f_n-f_m}{\Delta
                              W (\hbar \omega-\Delta W +i\gamma)} \nonumber \\
&=\frac{i(ev_F)^2\hbar}{2\pi^2l_b^2}\int\text{d}k_z\frac{1}{16}(1+\chi 
  k_z(\frac{2}{l_b^2}+k_z^2)^{-0.5})
  (1+\chi  k_z(\frac{4}{l_b^2}+k_z^2)^{-0.5}) \nonumber \\
&\times  \frac{f_n-f_m}{\hbar
  v_F(\sqrt{\frac{4}{l_b^2}+k_z^2}+\sqrt{\frac{2}{l_b^2}+k_z^2})
  (\hbar \omega-\hbar
  v_F(\sqrt{\frac{4}{l_b^2}+k_z^2}+\sqrt{\frac{2}{l_b^2}+k_z^2)
  }+i\gamma)} \nonumber \\
(\tilde{\sigma}_{LL})_{1\rightarrow 2}&=\frac{i(e
                              v_F)^2\hbar}{2\pi^2l_b^2}\int\text{d}k_z(C_{1}C_2v_{1}u_2)^2\frac{f_n-f_m}{\Delta
                              W (\hbar \omega-\Delta W +i\gamma)} \nonumber \\
&=\frac{i(ev_F)^2\hbar}{2\pi^2l_b^2}\int\text{d}k_z\frac{1}{16}(1-\chi 
  k_z(\frac{2}{l_b^2}+k_z^2)^{-0.5})
  (1+\chi  k_z(\frac{4}{l_b^2}+k_z^2)^{-0.5}) \nonumber \\
&\times  \frac{f_n-f_m}{\hbar v_F(\sqrt{\frac{4}{l_b^2}+k_z^2}-\sqrt{\frac{2}{l_b^2}+k_z^2}) (\hbar \omega-\hbar v_F(\sqrt{\frac{4}{l_b^2}+k_z^2}-\sqrt{\frac{2}{l_b^2}+k_z^2) }+i\gamma)} \nonumber 
\end{align}

Note that the occupation number $f_0$ of the $n = 0$ state depends on chirality even when Fermi energies are the same for both chiralities, since the energy  $W_0=-\chi \hbar v_F k_z$ depends on chirality.

\subsection{Longitudinal conductivity and plasmon dispersion for an arbitrary Fermi level}

Here we consider plasmons propagating along the magnetic field of a WSM with the Fermi level crossing an arbitrary number $N$ of Landau levels. The Hamiltonian is 
\begin{equation} 
\label{ham2} 
\mathcal{H}^{(\chi)} = \chi  v_F \mb{\sigma} \left( \mb{p}+\frac{e}{c}\mb{A}(\mb{r}) \right) - e \phi(\mb{r},t),
\end{equation}
where $\mb{A}(\mb{r})$ defines a constant magnetic field $\mb{B} \| \mb{z_0}$ whereas the electric field $\mb{E} \| \mb{z_0}$ of the plasmon is described by the scalar potential  $\phi = {\rm Re}\Phi e^{iqz -i\omega t}$. 

The kinetic equation for the electron distribution $f_{k_z}^{(n,\chi)}$ in the $n$th Landau subband is 
\be 
\label{kin}
\frac{\partial}{\partial t} f_{k_z}^{(n,\chi)} + \frac{\partial W_{n,\chi}}{\hbar \partial k_z} \frac{\partial}{\partial z} f_{k_z}^{(n,\chi)}  - eE \frac{\partial }{\hbar \partial k_z} f_{k_z}^{(n,\chi)} = 0.
\ee
Here we ignored relaxation. It can be added within the rate approximation as an imaginary part of frequency in the final expression for the conductivity. Note that for $n = 0$ 
$$ \frac{\partial W_{n=0,\chi}}{\hbar \partial k_z} = \chi v_F. $$
For $n \neq 0$ the presence of R- and L-fermions with opposite chiralities $\chi = \pm 1$ leads only to the degeneracy factor $g = 2$, as long as their Fermi energies are the same. 

Since we need the linear response, we linearize the distribution function in Eq.~(\ref{kin}) as $ f_{k_z}^{(n,\chi)} =  F_{k_z}^{(n,\chi)} + {\rm Re} \tilde{f}_{k_z}^{(n,\chi)} e^{iqz -i\omega t}$, which yields 
\be 
\label{flin} 
\tilde{f}_{k_z}^{(0,\chi)} = \frac{ieE}{(\omega - \chi q v_F)} \frac{\partial }{\hbar \partial k_z} F_{k_z}^{(0,\chi)}; \; 
\tilde{f}_{k_z}^{(n \neq 0,\chi)} = \frac{ieE}{(\omega - q \frac{\partial W_{n}}{\hbar \partial k_z} )} \frac{\partial }{\hbar \partial k_z} F_{k_z}^{(n \neq 0,\chi)}.
\ee

The complex amplitude of the current density is 
\begin{align}
\label{jz1} 
& j_z = {\rm Re} \tilde{j}_z e^{iqz -i\omega t}, \; \tilde{j}_z = \tilde{j}_0 + \sum_{n \neq 0} \tilde{j}_n;  \\
\label{jz2}
& \tilde{j}_0 = - \frac{e^2 B v_F}{4 \pi^2 \hbar^2 c} \int_{-\infty}^{\infty} \left( \tilde{f}_{k_z}^{(0,\chi = 1)} - \tilde{f}_{k_z}^{(0,\chi = -1)}  \right)\, dk_z, \\
\label{jz3}
& \tilde{j}_{n \neq 0}  = - 2 \frac{e^2 B}{4 \pi^2 \hbar^2 c} \int_{-\infty}^{\infty} \frac{\partial W_{n}}{\hbar \partial k_z}  \tilde{f}_{k_z}^{(n)}\,  dk_z.  
\end{align}
From Eq.~(\ref{jz2}) and the first of Eq.~(\ref{flin}) we obtain
\be 
\label{j0}
\tilde{j}_0 = - \frac{i e^3 B v_F}{4 \pi^2 \hbar^2 c} \left( \frac{1}{\omega - q v_F} \int_{-\infty}^{\infty} \frac{\partial }{ \partial k_z} F_{k_z}^{(0,\chi = 1)} \, dk_z - \frac{1}{\omega + q v_F} \int_{-\infty}^{\infty} \frac{\partial }{ \partial k_z} F_{k_z}^{(0,\chi = - 1)} \, dk_z \right) . 
\ee
Now we can derive the conductivity and the longitudinal component of the dielectric tensor by equating
\be 
\tilde{j}_0 + \sum_{n \neq 0} \tilde{j}_n = - \frac{i \omega (\epsilon_{zz} - \epsilon_b)}{4 \pi} E.
\ee
 Note that $F_{k_z \rightarrow - \infty}^{(0,\chi = 1)}\Rightarrow 1$, $F_{k_z \rightarrow  \infty}^{(0,\chi = 1)}\Rightarrow 0$, whereas $F_{k_z \rightarrow - \infty}^{(0,\chi = - 1)}\Rightarrow 0$ and $F_{k_z \rightarrow  \infty}^{(0,\chi = - 1)}\Rightarrow 1$. Therefore, Eq.~(\ref{j0}) yields
\be 
\label{j0-2}
\tilde{j}_0 =  \frac{i e^3 B v_F E}{2 \pi^2 \hbar^2 c}  \frac{\omega}{\omega^2 - q^2 v_F^2}  . 
\ee
The same result can be obtained from the quantum-mechanical density-matrix approach. We can similarly evaluate the integrals in Eq.~(\ref{jz3}) assuming the Fermi distribution with zero temperature. The result is 
\be
\label{ezz2}
\epsilon_{zz} = \epsilon_b - \omega_p^2 \left[ \frac{1}{\omega^2 - q^2 v_F^2} + 2 \sum_{n=1}^N \displaystyle  \frac{ \frac{1}{v_F} \left| \frac{\partial W_{n}}{\hbar \partial k_z}\right|_{W_n = E_F} }{ \omega^2 - q^2 \left( \frac{\partial W_{n}}{\hbar \partial k_z}\right)^2_{W_n = E_F} } \right].
\ee

The dispersion relation $\epsilon_{zz} = 0$ for a plasmon propagating along the magnetic field can be solved using Eq.~(\ref{ezz2}), leading to a cumbersome expression. In the limit of small $q$ one can put $q = 0$ in Eq.~(\ref{flin}) and replace 
$$ \tilde{f}_{k_z}^{(n\neq 0)} \approx \frac{ieE}{\omega}  \frac{\partial}{\hbar \partial k_z} F_{k_z}^{(n\neq 0)}. $$ 
This gives much simpler expressions: 
\be
\label{ezz3}
\epsilon_{zz} = \epsilon_b - \frac{\omega_p^2}{\omega^2} \left[ 1 + \frac{2}{v_F} \sum_{n=1}^N
\int_{-\infty}^{\infty} \left( \frac{\partial^2 W_{n}}{\hbar \partial k_z^2}\right) F_{k_z}^{(n)} \, dk_z \right], 
\ee
\be
\label{disp3}
\omega^2  =  \frac{\omega_p^2}{\epsilon_b} \left[ 1 + \frac{2}{v_F} \sum_{n=1}^N
\int_{-\infty}^{\infty} \left( \frac{\partial^2 W_{n}}{\hbar \partial k_z^2}\right) F_{k_z}^{(n)} \, dk_z \right]. 
\ee
These expressions are not limited to low temperatures and are valid for any unperturbed electron distribution $F_{k_z}^{(n)}$.

\end{document}